\documentclass[10pt,a4paper,onecolumn]{article}
\usepackage{marginnote}
\usepackage{graphicx}
\usepackage{xcolor}
\usepackage{authblk,etoolbox}
\usepackage{titlesec}
\usepackage{calc}
\usepackage{tikz}
\usepackage{hyperref}
\hypersetup{colorlinks,breaklinks,
            urlcolor=[rgb]{0.0, 0.5, 1.0},
            linkcolor=[rgb]{0.0, 0.5, 1.0}}
\usepackage{caption}
\usepackage{tcolorbox}
\usepackage{amssymb,amsmath}
\usepackage{ifxetex,ifluatex}
\usepackage{seqsplit}
\usepackage{fixltx2e} % provides \textsubscript
\usepackage{lmodern}
\usepackage[
  backend=biber,
]{biblatex}
\bibliography{paper/paper.bib}
\pdfoutput=1
\usepackage[top=3.5cm, bottom=3cm, right=1.5cm, left=1.0cm,
            headheight=2.2cm, reversemp, includemp, marginparwidth=4.5cm]{geometry}

\titleformat{\section}
  {\normalfont\sffamily\Large\bfseries}
  {}{0pt}{}
\titleformat{\subsection}
  {\normalfont\sffamily\large\bfseries}
  {}{0pt}{}
\titleformat{\subsubsection}
  {\normalfont\sffamily\bfseries}
  {}{0pt}{}
\titleformat*{\paragraph}
  {\sffamily\normalsize}

\usepackage{fancyhdr}
\makeatletter
\let\ps@plain\ps@fancy
\fancyheadoffset[L]{4.5cm}
\fancyfootoffset[L]{4.5cm}

\definecolor{linky}{rgb}{0.0, 0.5, 1.0}

\newtcolorbox{repobox}
   {colback=red, colframe=red!75!black,
     boxrule=0.5pt, arc=2pt, left=6pt, right=6pt, top=3pt, bottom=3pt}

\newcommand{\ExternalLink}{%
   \tikz[x=1.2ex, y=1.2ex, baseline=-0.05ex]{%
       \begin{scope}[x=1ex, y=1ex]
           \clip (-0.1,-0.1)
               --++ (-0, 1.2)
               --++ (0.6, 0)
               --++ (0, -0.6)
               --++ (0.6, 0)
               --++ (0, -1);
           \path[draw,
               line width = 0.5,
               rounded corners=0.5]
               (0,0) rectangle (1,1);
       \end{scope}
       \path[draw, line width = 0.5] (0.5, 0.5)
           -- (1, 1);
       \path[draw, line width = 0.5] (0.6, 1)
           -- (1, 1) -- (1, 0.6);
       }
   }

\patchcmd{\@maketitle}{center}{flushleft}{}{}
\patchcmd{\@maketitle}{center}{flushleft}{}{}
\patchcmd{\@maketitle}{\LARGE}{\LARGE\sffamily}{}{}
\def\maketitle{{%
  
  \AB@maketitle}}
\makeatletter
\renewcommand\AB@affilsepx{ \protect\Affilfont}
\renewcommand\AB@affilnote[1]{{\bfseries #1}\hspace{3pt}}
\makeatother

\renewcommand\Affilfont{\sffamily\small\mdseries}
\ifnum 0\ifxetex 1\fi\ifluatex 1\fi=0 % if pdftex
  \usepackage[T1]{fontenc}
  \usepackage[utf8]{inputenc}

\else % if luatex or xelatex
  \ifxetex
    \usepackage{mathspec}
  \else
    \usepackage{fontspec}
  \fi
  \defaultfontfeatures{Ligatures=TeX,Scale=MatchLowercase}

\fi
\IfFileExists{upquote.sty}{\usepackage{upquote}}{}
\IfFileExists{microtype.sty}{%
\usepackage{microtype}
\UseMicrotypeSet[protrusion]{basicmath} % disable protrusion for tt fonts
}{}

\usepackage{hyperref}
\hypersetup{unicode=true,
            pdftitle={AstronomicAL: an interactive dashboard for visualisation, integration and classification of data with Active Learning},
            pdfborder={0 0 0},
            breaklinks=true}
\usepackage{graphicx,grffile}
\makeatletter
\def\maxwidth{\ifdim\Gin@nat@width>\linewidth\linewidth\else\Gin@nat@width\fi}
\def\maxheight{\ifdim\Gin@nat@height>\textheight\textheight\else\Gin@nat@height\fi}
\makeatother
\setkeys{Gin}{width=\maxwidth,height=\maxheight,keepaspectratio}
\IfFileExists{parskip.sty}{%
\usepackage{parskip}
}{% else
\setlength{\parindent}{0pt}
\setlength{\parskip}{6pt plus 2pt minus 1pt}
}
\ifx\paragraph\undefined\else
\let\oldparagraph\paragraph
\renewcommand{\paragraph}[1]{\oldparagraph{#1}\mbox{}}
\fi
\ifx\subparagraph\undefined\else
\let\oldsubparagraph\subparagraph
\renewcommand{\subparagraph}[1]{\oldsubparagraph{#1}\mbox{}}
\fi

\title{AstronomicAL: an interactive dashboard for visualisation, integration and classification of data with Active Learning}

        \author[1]{Grant Stevens}
        \author[2]{Sotiria Fotopoulou}
        \author[2]{Malcolm N. Bremer}
        \author[1]{Oliver Ray}

      \affil[1]{Department of Computer Science, Merchant Venturers Building, University of Bristol, Woodland Road, Bristol, BS8 1UB}
      \affil[2]{School of Physics, HH Wills Physics Laboratory, University of Bristol, Tyndall Avenue, Bristol, BS8 1TL}

  \date{\vspace{-5ex}}

\begin{document}
\maketitle

\marginpar{
  %\hrule
  \sffamily\small

  {\bfseries DOI:} \href{https://doi.org/10.21105/joss.03635}{\color{linky}{10.21105/joss.03635}}

  \vspace{2mm}

  {\bfseries Software}
  \begin{itemize}
    \setlength\itemsep{0em}
    \item \href{https://github.com/openjournals/joss-reviews/issues/3635}{\color{linky}{Review}} \ExternalLink
    \item \href{https://github.com/grant-m-s/astronomicAL}{\color{linky}{Repository}} \ExternalLink
    \item \href{https://zenodo.org/record/5396671}{\color{linky}{Archive}} \ExternalLink
  \end{itemize}

  \vspace{2mm}

  {\bfseries Submitted:} 06 August 2021\\
  {\bfseries Published:} 03 September 2021

  \vspace{2mm}
  {\bfseries Licence}\\
  Authors of papers retain copyright and release the work under a Creative Commons Attribution 4.0 International License (\href{http://creativecommons.org/licenses/by/4.0/}{\color{linky}{CC-BY}}).
}

\hypertarget{summary}{%
\section{Summary}\label{summary}}

AstronomicAL is a human-in-the-loop interactive labelling and training dashboard that allows users to create reliable datasets and robust classifiers using active learning. This technique prioritises data that offer high information gain, leading to improved performance using substantially less data. The system allows users to visualise and integrate data from different sources and deal with incorrect or missing labels and imbalanced class sizes. AstronomicAL enables experts to visualise domain-specific plots and key information relating both to broader context and details of a point of interest drawn from a variety of data sources, ensuring reliable labels. In addition, AstronomicAL provides functionality to explore all aspects of the training process, including custom models and query strategies. This makes the software a tool for experimenting with both domain-specific classifications and more general-purpose machine learning strategies. We illustrate using the system with an astronomical dataset due to the field’s immediate need; however, AstronomicAL has been designed for datasets from any discipline. Finally, by exporting a simple configuration file, entire layouts, models, and assigned labels can be shared with the community. This allows for complete transparency and ensures that the process of reproducing results is effortless.

\clearpage

\begin{figure}[!ht]
  \includegraphics[width=0.86\textwidth]{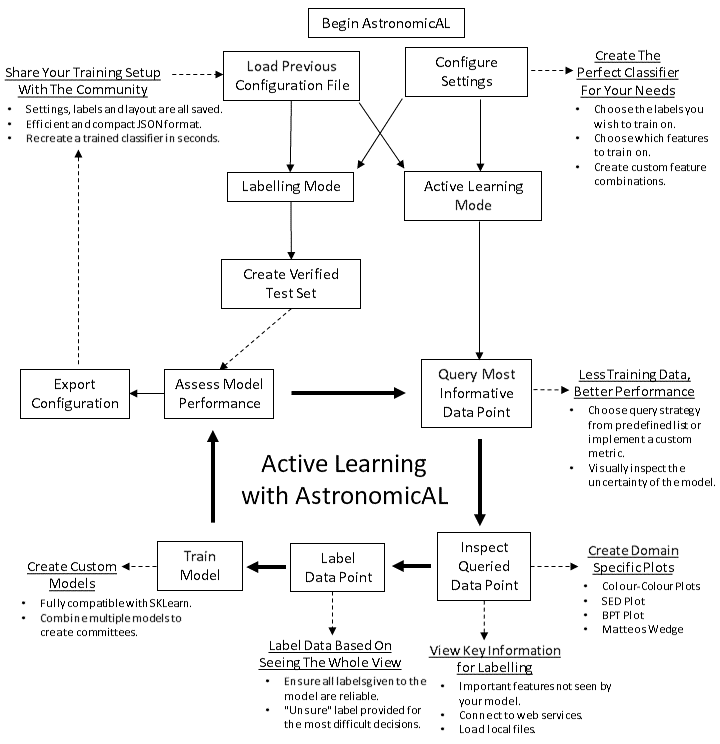}
  \caption{AstronomicAL workflow from the perspective of an astronomy user. Each part of the workflow improves the reliability of labels, leading to enhanced performance of any classifiers produced. AstronomicAL allows the user to tailor this workflow specifically to their domain with its modular and extensible design.}
  \label{fig:workflow}
\end{figure}

\hypertarget{statement_of_need}{%
\section{Statement of Need}\label{statement_of_need}}

Active learning has proven to be an effective method in machine learning over the last few decades \cite{settles2008analysis, baldridge2009well, gal2017deep}. In more recent years, there has been a growing adoption in all domains of scientific research \cite{walmsley2020galaxy, eisenstein2020active}. This is evident by the increasing popularity of machine learning frameworks offering an active learning element\footnote{https://github.com/topics/active-learning}. However, the purpose of these frameworks is to allow users to integrate active learning into their machine learning workflows. Any additional functionality such as data exploration and interactive labelling is left to the user to implement. Due to each domain's varying requirements, researchers will often need multiple specialised but disjoint tools that hinder effective and efficient interactive labelling. When data from multiple sources is required, presenting all the information in one system is essential. Without a tool that combines these key features, a significant barrier exists to the adoption and effectiveness of active learning in research. AstronomicAL's goal is to remove this barrier to allow more researchers to access the benefits of active learning.

\clearpage

Active learning \cite{settles2012active} removes the requirement for large amounts of labelled training data whilst still producing high accuracy models. This is extremely important as with ever-growing datasets; it is becoming impossible to manually inspect and verify ground truth used to train machine learning systems. The reliability of the training data limits the performance of any supervised learning model, so consistent classifications become more problematic as data sizes increase. The problem is exacerbated when a dataset does not contain any labelled data, preventing supervised learning techniques entirely. AstronomicAL has been developed to tackle these issues head-on and provide a solution for any large scientific dataset.

It is common for active learning to query areas of high uncertainty; these are often in the boundaries between classes where the expert’s knowledge is required. To facilitate this human-in-the-loop process, AstronomicAL provides users with the functionality to fully explore each data point chosen. This allows them to inject their domain expertise directly into the training process, ensuring that assigned labels are both accurate and reliable.

AstronomicAL has been extensively validated on astronomy datasets. These are highly representative of the issues that we anticipate will be found in other domains for which the tool is designed to be easily customisable. Such issues include the volume of data (millions of sources per survey), vastly imbalanced classes and ambiguous class definitions leading to inconsistent labelling. AstronomicAL has been developed to be sufficiently general for any tabular data and can be customised for any domain. For example, we provide the functionality for data fusion of catalogued data and online cutout services for astronomical datasets.

Using its modular and extensible design, researchers can quickly adapt AstronomicAL for their research to allow for domain-specific plots, novel query strategies, and improved models. Furthermore, there is no requirement to be well-versed in the underlying libraries that the software uses. This is due to large parts of the complexity being abstracted whilst allowing more experienced users to access full customisability.

As the software runs entirely locally on the user’s system, AstronomicAL provides a private space to experiment whilst providing a public mechanism to share results. By sharing only the configuration file, users remain in charge of distributing their potentially sensitive data, enabling collaboration whilst respecting privacy. The full workflow for AstronomicAL is shown in \autoref{fig:workflow}.

\clearpage

\begin{figure}[!ht]
  \includegraphics{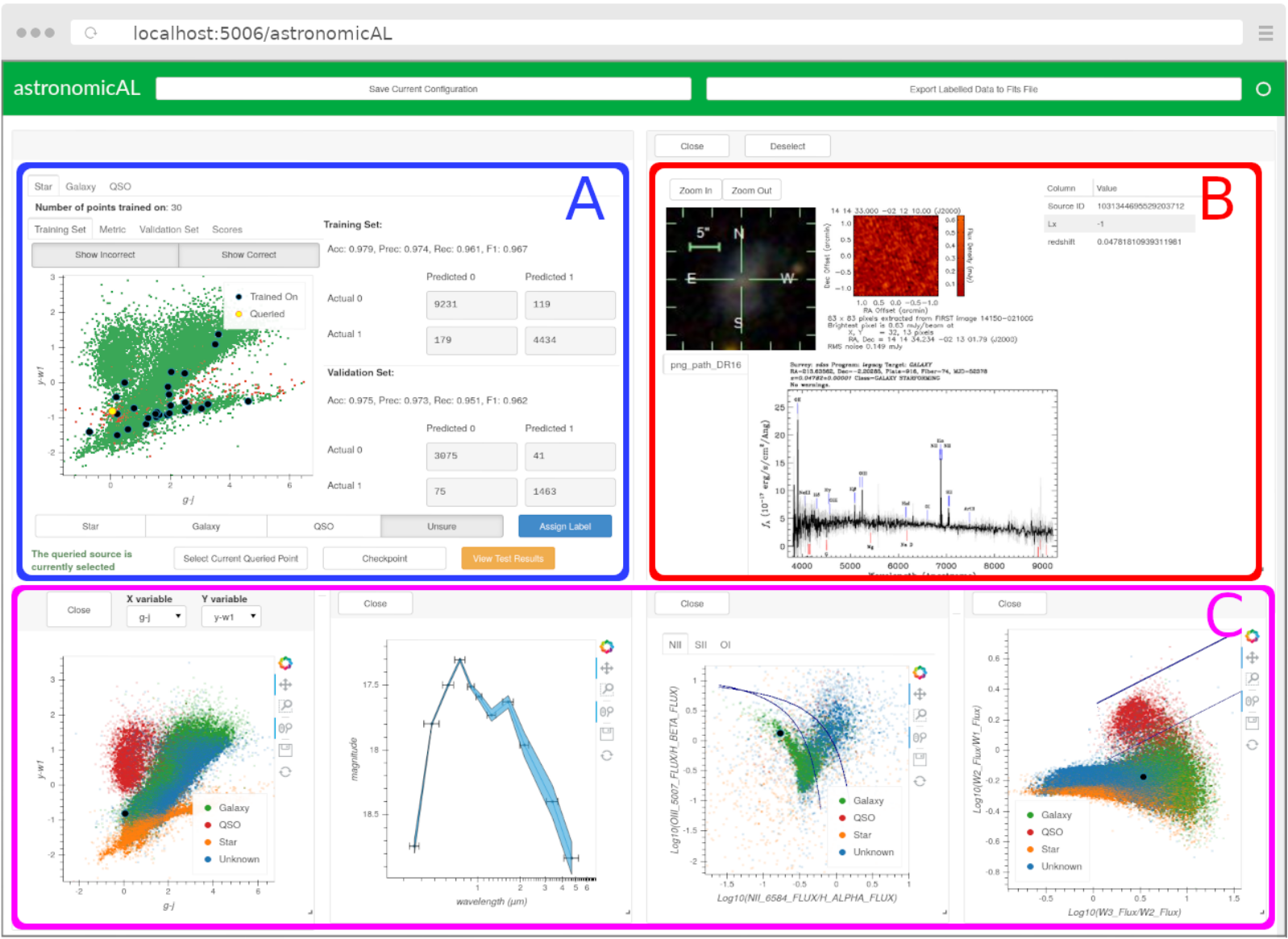}
  \caption{The dashboard is set up in 3 main areas. A) Active learning and labelling panel which controls the machine learning functionality. Users can train separate models for each class and see how each performs as additional points are added to the model. B) User-defined information shown for the currently active data point, such as specific columns, local images and online cutout services. C) Customisable domain-specific plots that can render millions of data points.}
  \label{fig:full_layout}
\end{figure}

\hypertarget{active_learning}{%
\section{Active Learning and Classification}\label{active_learning}}

Active learning utilises methods for predicting which part of the search space of unseen examples would be most likely to improve classification accuracy. By querying these key examples and presenting them directly to the user for inspection and labelling, active learning removes the reliance on non-verified data. In the optimal case, it provides users with simple classifications that have a high impact on model performance, leading to a dramatic reduction in the number of data points required for training a classifier.

By making use of the modAL \cite{modAL2018} active learning framework, AstronomicAL allows users to take full advantage of active learning techniques, leading to improved classifiers. Scikit-Learn \cite{scikit-learn} models are fully supported in AstronomicAL, and the software will be extended to support Pytorch \cite{pytorch} and Tensorflow \cite{tensorflow2015-whitepaper} in the future.

Astronomy datasets are frequently the result of the fusion of multiple heterogeneous data sources. Often in machine learning, only information that is available across all data points is usable during training. However, in astronomy, where \textit{any} available information could be critical in determining an object’s true class, it must be used whenever this information is available. With the flexibility to choose which information is essential for classification, AstronomicAL ensures that users are presented with this critical information whenever available - directly leading to improved confidence and justifiability for any assigned labels. Although this extra information remains unseen by the model, by using AstronomicAL, we inject this information into the training process by labelling the data based on all the available information rather than just the features the model sees.

\begin{figure}[!ht]
  \centering
  \includegraphics[width=0.80\textwidth]{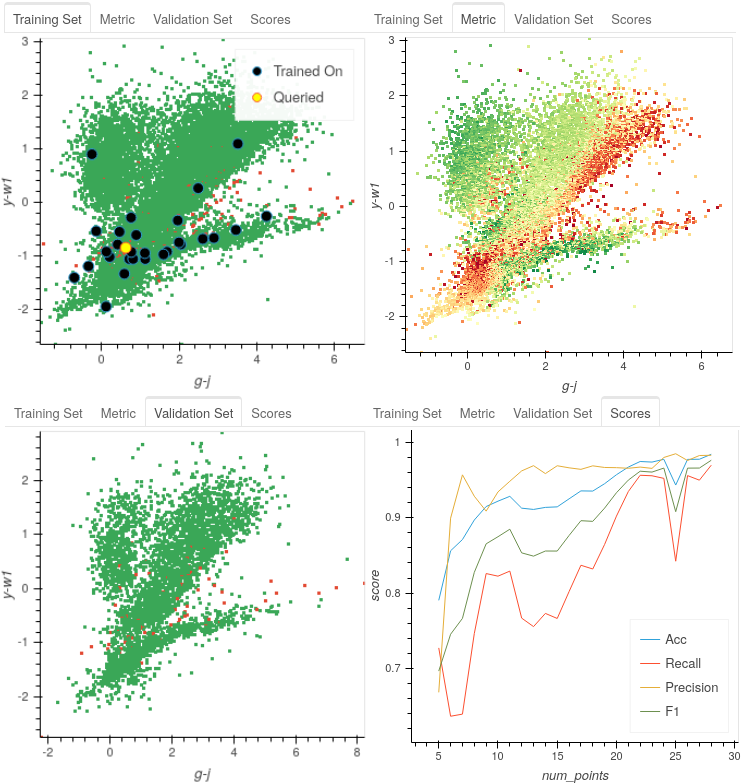}
  \caption{AstronomicAL allows users to view all key information during training. Top Left: Training set showing points that have been trained on, current queried point and correct and incorrect predictions. Top Right: Visualisation of the model’s confidence of each data point. Bottom Left: Validation set showing correct and incorrect predictions of the model. Bottom Right: The performance scores of the model as new points are added to the training set.}
  \label{fig:active_learning}
\end{figure}

In \autoref{fig:full_layout} we show an example of a classification task using an extract of the astronomical dataset used in \textcite{Fotopoulou2018CPz}. \autoref{fig:full_layout} (A) shows the benefits of active learning where with only 30 data points given as training data to the model, we can classify Stars with 97.5\% accuracy. \autoref{fig:full_layout} (B) shows the extra information chosen by the user to improve confidence in the assigned labels. This information can be retrieved from web services or local files. In this example, we are showing the SDSS\footnote{http://skyserver.sdss.org/dr16/en/help/docs/api.aspx\#imgcutout} \cite{Ahumada2020DR16} and FIRST\footnote{https://third.ucllnl.org/cgi-bin/firstcutout} \cite{Becker1995Radio} cutout services. \autoref{fig:full_layout} (C) shows how domain-specific plots can be created and visualised to gain extra insight into the data, aiding the researcher in their ability and confidence in labelling data points. \autoref{fig:active_learning} shows the model performance plots available to the user for each classifier being trained. These include training and validation correctness plots, tracking of performance scores as new data points are added, and a visualisation of the model's confidence of each data point.

\clearpage

\hypertarget{interactivity}{%
\section{Interactivity}\label{interactivity}}

Using the Panel dashboard library \cite{philipp_rudiger_panel_2021_4573728}, paired with Holoviews \cite{philipp_rudiger_holoviews_2021_4581995}, Bokeh \cite{Bokeh2018} and Datashader \cite{james_a_bednar_datashader_2020_3987379} visualisation libraries, AstronomicAL provides users with interactive plots, enabling zooming and panning of millions of data points in real-time whilst also giving the ability to rearrange and resize plots dynamically to optimise screen layout.

Plots are also interconnected, with any updates projected to all panels simultaneously, allowing analysis of how the properties of a particular point fit in the context of the whole dataset. This enables researchers to gain insight and identify trends - crucial for reliable labelling.

\begin{figure}
  \centering
  \includegraphics[width=0.85\textwidth]{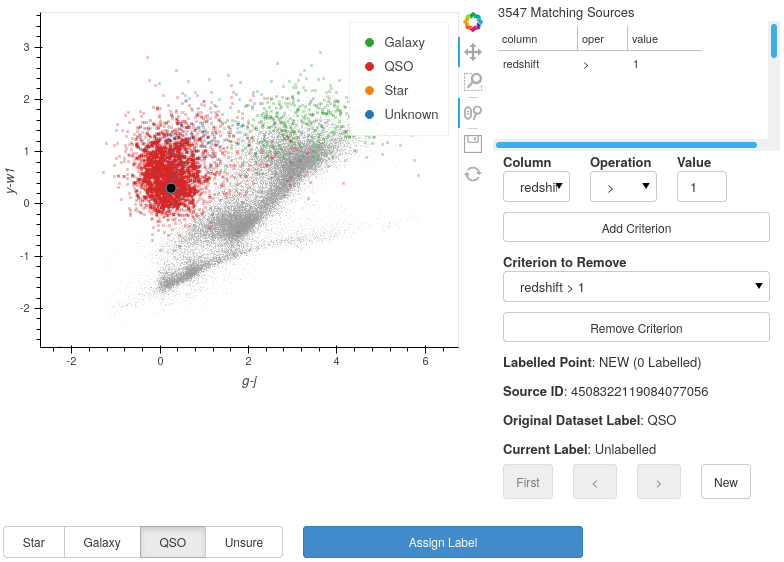}
  \caption{AstronomicAL enables the creation of curated test sets. The sample is selected according to user-defined criteria and verified visually. This test set can then be used during the active learning training process to ensure that the model will be sufficiently generalisable to future data.}
  \label{fig:labelling_mode}
\end{figure}

The interactive labelling functionality is not limited to only the training stage; as shown in \autoref{fig:labelling_mode}, users can curate a labelled test set to sufficiently demonstrate the validity and generalisability of their model. Furthermore, to facilitate and encourage only assigning labels that the user trusts, AstronomicAL allows users to mark any example as \textit{unsure}. Such examples are excluded from the training set, ensuring that all training data are of high quality.

In summary, all design decisions have been in response to user feedback and were explicitly tailored to improve the visualisation of data, irrespective of the data’s specific nature, to ensure that AstronomicAL is a tool for any discipline.

\hypertarget{acknowledgements}{%
\section{Acknowledgements}\label{acknowledgements}}

Grant Stevens acknowledges financial support from the UKRI for a Centre for Doctoral Training studentship in Interactive Artificial Intelligence at the University of Bristol. (EP/S022937/1)

\printbibliography

\end{document}